\documentstyle[aps,psfig,tighten,floats]{revtex}

\newcommand{\SC}{\scriptscriptstyle} 
\newcommand{\SS}{\scriptstyle} 

\begin{document}

\title{\bf 3+1 spinfoam model of quantum gravity 
with spacelike and timelike components}

\author{Alejandro Perez and Carlo Rovelli \\ {\it Centre de 
Physique
Th\'eorique - CNRS, Case  907, Luminy,
             F-13288 Marseille, France}, and \\
{\it Physics Department, University of Pittsburgh, 
             Pittsburgh, Pa 15260, USA}}

\maketitle

\begin{abstract}

We present a spinfoam formulation of Lorentzian quantum General
Relativity.  The theory is based on a simple generalization of an
Euclidean model defined in terms of a field theory over a group.  The
model is an extension of a recently introduced Lorentzian model, in
which both timelike and spacelike components are included.  The
spinfoams in the model, corresponding to quantized 4-geometries, carry
a natural non-perturbative local causal structure induced by the
geometry of the algebra of the internal gauge ($sl(2,C)$).  Amplitudes
can be expressed as integrals over the spacelike unit-vectors
hyperboloid in Minkowski space, or the imaginary Lobachevskian space.

\end{abstract}
 
\section{Introduction}

Most of the work in non-perturbative quantum gravity is restricted to
the unphysical Euclidean sector, deferring the construction of the
physical Lorentzian theory.  A step towards the Lorentzian theory has
been recently taken in Refs.\,\cite{BC2,lac}, with the definition of
spinfoam models based on $SL(2,C)$ representation theory.  However,
these models includes only (simple) representations in the continuous
series, and not the ones in the discrete series.  Since the signature
of the surfaces in the spinfoam is determined by the sign of the
Casimir, which is opposite in the two series, all surfaces of the
model turn out to have the same signature.  In a more realistic model,
on the other hand, we expect spacelike as well as timelike surfaces to
appear, and thus all (simple) representation to contribute.  In this
paper, we introduce a new model, which includes both kinds of
representations.  As a consequence, the model incorporates a
combinatorial non-perturbative notion of local causality associated to
quantum spacetime.  Surfaces in a given spinfoam can be classified as
timelike or spacelike, according to the kind of simple representations
by which they are colored.

We recall that spinfoam models provide a framework for background
independent diffeomorphism-invariant quantum field theory and quantum
gravity in particular \cite{Reisenberger,Iwasaki,Baez,rr,Roberto,KF,BC}. 
They provide a rigorous implementation of the
Wheeler-Misner-Hawking\cite{misner,haw} sum over geometries
formulation of quantum gravity.  The 4-geometries summed over are
represented by spinfoams.  These are defined as colored 2-complexes. 
A 2-complex $J$ is a (combinatorial) set of elements called
``vertices'' $v$, ``edges'' $e$ and ``faces'' $f$, and a boundary
relation among these.  A spinfoam is a 2-complex plus a ``coloring''
$N$, that is an assignment of an irreducible representation $N_{f}$ of
a given group $G$ to each face $f$ and of an intertwiner $i_{e}$ to
each edge $e$.  The model is defined by the partition function
\begin{equation}
Z = \sum_{J}\ {\cal N}(J) \sum_{N}\ \prod_{f\in J} 
A_{f}(N_{f}) \ \prod_{e\in J} A_{e}(N_e) \prod_{v\in
J} A_v(N_v),  \label{Z}
\end{equation}
where $A_{f}$, $A_{e}$ and  $A_v$ correspond to the amplitude associated 
to faces, edges,  and vertices respectively (they are given functions of the 
corresponding colors). ${\cal N}(J)$ is a normalization factor
for each 2-complex. 

Spinfoam models can be obtained as the perturbative expansion of a
field theory over a group manifold\cite{cm}.  In this language,
spinfoams appear as Feynman diagrams of a scalar field theory over a
group.  More precisely equation (\ref{Z}) can be obtained by the
perturbative expansion of
\begin{equation}
Z=\int {\cal D}\phi\; e^{-S[\phi]}
\end{equation}
in momentum space.  In particular, the topological models of BF theory
can be described in this framework
\cite{t,Ooguri:1992b,CraneYetter,CraneYetter1}.  The field theory
approach has several advantages.  It implements automatically the sum
over 2-complexes $J$, and in particular, fixes the ${\cal N}(J)$ value
in (\ref{Z}).  This sum restores full general covariance of a theory
with local degrees of freedom such as general relativity (GR)
\cite{Baez,dfkr}.  Other advantages of the field theory formulation
are related to the possibility of performing formal manipulations
directly on configuration space of the field theory, which turn out to
be much simpler than working directly with the state sum model of
equation (\ref{Z}) (i.e., momentum space).  Examples of this are the
proof of topological invariance for the BF models due to
Ooguri\cite{Ooguri:1992b}, and the definition of the Lorentzian model
mentioned above.  In this last case, the non-compactness of $SL(2,C)$
introduces additional complications that are easier to deal with in
the field theory formulation.

Spinfoam models related to gravity have been obtained as modifications
of topological quantum field theories by implementing the constraint
that reduce BF theory to GR \cite{BC,Reisenberg97,iwa0,ac}. 
Essentially this constraint amounts to restrict to simple
representations only.  The resulting models correspond to a
diffeomorpism invariant lattice-like quantization of Plebanski's
formulation of GR \cite{cuate}.  In the field theory framework, the
constraints reducing BF theory to GR are naturally implemented by
imposing a symmetry requirement on the field action defining the BF
model \cite{dfkr}.  In the Euclidean models, the field is defined over
$SO(4)$, and the BF to GR constraint is imposed as a symmetry under an
$SO(3)$ subgroup.  Amplitudes turn out to be expressed as integrals
over the homogeneous space $S_{3}=SO(4)/SO(3)$ and the theory is
controlled by the harmonic analysis on $S_{3}$, which contains the
simple representations of $SO(4)$.  In the Lorentzian models
introduced in Refs.\,\cite{BC2,lac}, the field is defined over
$SL(2,C)$, and the BF to GR constraint is imposed as a symmetry under
an $SU(2)$ subgroup.  Amplitudes turn out to be expressed as integrals
over the homogeneous space $SL(2,C)/SU(2)$, which is the (real)
Lobachevskian space: the upper hyperboloid of unit norm timelike
vectors in in Minkowski space.  The theory is controlled by the
harmonic analysis on this space, which contains the continuous simple
representations of $SL(2,C)$ only.

In the model we introduce here, the field is still defined over
$SL(2,C)$, but the BF to GR constraint is imposed as a symmetry under
an $SU(1,1)\times Z_{2}$ subgroup.  Amplitudes turn out to be
expressed as integrals over the homogeneous space
$SL(2,C)/(SU(1,1)\times Z_{2})$, which is the {\em imaginary\/}
Lobachevskian space: the hyperboloid of unit norm spacelike vectors in
in Minkowski space, with opposite points identified.  The theory is
controlled by the harmonic analysis on this space, which contains
simple representations of $SL(2,C)$ in the continuous as well as in
the discrete series.  Thus, we obtain a model that implements the BF
to GR constraints and in which the quantum four geometries summed over
have spacelike as well as timelike surfaces.

The paper is organized as follows.  In the next section we describe
the general setting in which the Lorentzian spinfoam model of
reference \cite{lac} was defined.  The new model is defined in the
same framework in terms of a simple modification of the one in
\cite{lac}.  Both variants are presented in the corresponding
subsections.  We obtain the edge and vertex amplitudes of the models
as integrals over the Lobachevskian spaces of a kernel $K^{\SC\pm}$. 
In section \ref{app2} we compute these kernels for the two models.  In
the appendix we present a compendium of known results on harmonic
analysis and representation theory of $SL(2,C)$ on which our
construction is based. 

\section{$SL(2,C)$ state sum models of Lorentzian QG}

In this section we briefly discuss the general framework
of spinfoam models defined as a field theory over a group.
In particular we review the derivation of the Lorentzian 
spinfoam model of \cite{lac}. The implementation of the simplicity
constraints is encoded the symmetries of the interaction. 
Those symmetries are implemented in the field action by means of 
group averaging techniques. The new model differs from its previous 
relative in the kind of 
symmetries required (different implementation of the constraints).
For that reason we start the section with the simultaneous treatment
of both theories since the formal manipulation necessary to derive the model
are the same in each one of them. In the two subsection that follow this 
general derivation we explicitly state the results for both models.   

We start with a field $\phi(g_1,g_2,g_3,g_4)$ over 
${ SL(2,C)\times SL(2,C)\times SL(2,C)\times SL(2,C)}$.  We
assume the field has compact support and is symmetric under arbitrary
permutations of its arguments.
\footnote{ This symmetry guarantees
arbitrary 2-complexes $J$ to appear in the Feynman expansion\cite{dfkr}.}
We define the projectors $P_{\gamma}$ and $P^{\SC (\pm)}_{u}$ as
\begin{equation} 
    \label{pg} P_{g}\phi(g_i)
\equiv \int d\gamma  \ {\phi}(g_i\gamma),
\end{equation} 
and
\begin{equation} \label{pu}        
P^{\SC (\pm)}_{u}\phi(g_i) \equiv \int_{U^{\SC (\pm)}} du^{\SC (\pm)}_i \ {\phi}( g_i u_{i}),
\end{equation}
with $\gamma \in SL(2,C)$, and $u_i \in U^{\SC (\pm)} \subset SL(2,C)$, where $U^{\SC (\pm)}$ correspond
the $SL(2,C)$ subgroups defined as follows. Think of the
vector representation of $SL(2,C)$, i.e., four dimensional irreducible representation
defined on $\Re^4$. The $SL(2,C)$ action defines a Minkowski metric in $\Re^4$.
Let it be of signature $(+,-,-,-)$. Consider the time-like line $(\lambda,0,0,0)$ and the
space-like line $(0,0,0,\lambda)$ for $\lambda \in \Re$, and associate to them the one parameter 
family of $2\times 2$ matrices ${\ell}^{\SC +}=\lambda \sigma_0$ and ${\ell}^{\SC -}=\lambda \sigma_3$ 
respectively ($\sigma_{\mu}$ being the identity matrix for $\mu=0$ and the Pauli matrices for $\mu=1,2,3$). 
We define the subgroup $ U^{\SC (\pm)} \subset SL(2,C)$ as the subgroup 
leaving invariant $\ell^{\SC \pm}$ respectively under the action $\ell \rightarrow u\ell u^{\dagger}$.
Clearly  $U^{\SC (+)}$ is isomorphic to $SU(2)$. $U^{\SC (-)}$, on the other hand, is 
isomorphic to  $SU(1,1) \times Z_2$ %
\footnote{
The elements of $Z_2$ can be realized as the $2 \times 2$ 
matrices $\left[\begin{array}{c} 1 \ 0 \\ 0 \ 1 
\end{array}\right]$, and
$\left[\begin{array}{c} 0 \ i \\ i \ 0 \end{array} \right]$ respectively.}.
Finally, $d\gamma$ and $du^{\SC (\pm)}$ denote the corresponding invariant measures.  

We define the following two actions which give rise to the
spinfoam models considered in the paper
\begin{equation}
    \label{action} 
S^{\SC (\pm)}[\phi]=\int dg_i \left[ P_{\gamma} \phi(g_i) \right]^2 + 
{1\over 5!} \int dg_i \left[ P_{\gamma} P^{\SC (\pm)}_{u} \phi(g_i) \right]^5,
\end{equation} 
where $\gamma_{i} \in SL(2,C)$, $\phi(g_i)$ denotes
$\phi(g_1,g_2,g_3,g_4)$, and the fifth power in the interaction term
is notation for 
\begin{equation}
 \left[\phi(g_i)\right]^5:=\phi(g_1,g_2,g_3,g_4)\
 \phi(g_4,g_5,g_6,g_7)\ \phi(g_7,g_3,g_8,g_9)\
 \phi(g_9,g_6,g_2,g_{10}) \ \phi(g_{10},g_8,g_5,g_1).
\end{equation}
The $\gamma$ integration projects the field into the space of gauge
invariant fields, namely, those such that $\phi(g_i)=\phi(g_i\mu)$ for
$\mu \in SL(2,C)$.%
\footnote{Because of this gauge invariance, the action (\ref{action})
is proportional to the trivial diverging factor $\int d\gamma$.  This
divergence could be fixed easily, for instance by gauge fixing and
just dropping one of the group integrations.  For the clarity of the
presentation, however, we have preferred to keep gauge invariance
manifest, use the action formally to generate the Feynman expansion,
and drop the redundant group integration whenever needed.\label{foot}}
We continue the construction of the spinfoam models
corresponding to $S^{(+)}[\phi]$, and $S^{-}[\phi]$ in general and we particularize to each
case in the following two subsections.
The vertex and propagator of the theories are simply given by a set of
delta functions on the group, as illustrated in \cite{ac}, to which we
refer for details.  Feynman diagrams correspond to arbitrary 2-complex
$J$ with 4-valent edges (bounding four faces), and 5-valent vertices
(bounding five edges).  Once the configuration variables $g_i$ are
integrated over, the Feynman amplitudes reduce to integrals over the
group variables $\gamma$ and $u$ in the projectors in (\ref{action}). 
These end up combined as arguments of one delta functions per face
\cite{ac}.  That is, a straightforward computation yields
\begin{equation} 
    \label{mart}A^{(\pm)}(J)=\int_{U^{(\pm)}} du^{\SC (\pm)}d\gamma \prod_{{e }}
\prod_{{f }} \delta(\gamma^{\SC(1)}_{e_{1}}u^{\SC (\pm)}_{\SC 
1f}\gamma^{\SC(2)}_{e_{1}}u^{\SC \prime (\pm) }_{\SC
1f}\gamma^{\SC(3)}_{e_{1}} \dots \gamma^{\SC(1)}_{e_{N}}u^{\SC (\pm)}_{\SC 
Nf}\gamma^{\SC(2)}_{e_{N}}u^{\SC \prime (\pm) }_{\SC
Nf}\gamma^{\SC(3)}_{e_{N}}),
\end{equation} 
where $e$ and $f$ denote the set of edges and faces of the corresponding 2-complex $J$.
In this equation, $\gamma^{\SC(1)}_{e_{}}$, and
$\gamma^{\SC(3)}_{e_{}}$ come from the group integration in the
projectors $P_{\gamma}$ in the two vertices bounding the edge $e$. 
$\gamma^{\SC(2)}_{e_{}}$ comes from the projector $P_{\gamma}$ in the
propagator defining the edge $e$.  Finally, $u^{\SC (\pm)}_{\SC  1f}$ and
$u^{\SC \prime (\pm) }_{\SC 1f}$ are the integration variables in
the projector $P_{h}$ in the two vertices.  Notice that each $u$
integration variable appears only once in the integrand, while each
$\gamma$ integration variable appears in four different delta's (each
edge bounds four faces).  The index $N$ denotes the number of edges of
the corresponding face.  Now we use equation (\ref{vani}) to expand
the delta functions in terms of irreducible representations of
$SL(2,C)$. Only the representations $(n,\rho)$ in the principal series
contribute to this expansion. We obtain 
\begin{equation}
    \label{aj}
A^{\SC (\pm)}(J)=\sum_{n} \int_{\rho_f} d\rho \prod_{{f }} (\rho_f^2+n_f^2) \int
\prod_{{e }} d\gamma du^{\SC (\pm)}\ {\rm Tr}\left[ \bar D^{n_f
\rho_f}(\gamma^{\SC(1)}_{e_{1}}u^{\SC (\pm)}_{\SC 
1f}\gamma^{\SC(2)}_{e_{1}}u^{\SC \prime (\pm) }_{\SC
1f}\gamma^{\SC(3)}_{e_{1}} \dots \gamma^{\SC(1)}_{e_{N}}u^{\SC (\pm)}_{\SC 
Nf}\gamma^{\SC(2)}_{e_{N}}u^{\SC \prime (\pm) }_{\SC
Nf}\gamma^{\SC(3)}_{e_{N}} )\right].  
\end{equation} 
Next, we rewrite this equation in terms of the matrix elements 
$\bar D^{n \rho}_{j_1 q_1 j_2 q_2}(\gamma)$ of the representation $(n,\rho)$ 
in the canonical basis, defined in the
appendix.  The trace becomes
\begin{eqnarray}
\nonumber && {\rm Tr}\left[
\bar D^{n_f \rho_f}(\gamma^{\SC(1)}_{e_{1}}u^{\SC (\pm)}_{\SC 
1f}\gamma^{\SC(2)}_{e_{1}}u^{\SC \prime (\pm) }_{\SC 1f}\gamma^{\SC(3)}_{e_{1}}
\dots \gamma^{\SC(1)}_{e_{N}}u^{\SC (\pm)}_{\SC  Nf}\gamma^{\SC(2)}_{e_{N}}u^{\SC
\prime (\pm)}_{\SC Nf}\gamma^{\SC(3)}_{e_{N}}  )\right] =  \\ 
&&\ \ \ \ \ \ \ \ \ \ \ \ \ \ \ \ \ \ \ \ \ \ \ \ \ \ \ \ \ \ \ \ \ \ \ \ \ 
\ \ \ \ \ \ \ \ \ \ \ \ \bar D^{n_f \rho_f}_{j_1 q_1 j_2 q_2}(\gamma^{\SC(1)}_{e_{1}}) 
\bar D^{n_f \rho_f}_{j_2 q_2 j_3 q_3}(u^{\SC (\pm)}_{\SC 1f}) 
\bar D^{n_f \rho_f}_{j_3 q_3 j_4 q_4}(\gamma^{\SC(2)}_{e_{1}}) 
\dots \bar D^{n_f \rho_f}_{j_{.} q_{.} j_1 q_1}(\gamma^{\SC(3)}_{e_{N}}). 
\end{eqnarray}
Repeated indices are summed, and the range of the $j_{n}$ and $q_{n}$
indices is specified in the appendix. 
The integration of $\bar D^{n_f \rho_f}(u^{\SC (\pm)}_{\SC 1f})_{j_2 q_2 j_3 q_3}$ over  
$u^{\SC (\pm)}_{\SC 1f}$ is zero if in the corresponding representation there are 
no invariant vectors under the action of $U^{\SC (\pm)}$.
\footnote{This projection implements the constraint that reduces BF
theory to GR. Indeed, the generators of $SL(2,C)$ are identified with
the classical two-form field $B$ of BF theory.  The generators of the
simple representations satisfy precisely the BF to GR constraint. 
Namely $B$ has the appropriate $e\wedge e$ form \cite{BC,Baez}.}
It can be written as
\begin{equation}\label{wpm}
\int_{U^{\SC (\pm)}} du^{\SC (\pm)} \bar D^{n \rho}(u^{\SC (\pm)})_{j_1 q_1 j_2 q_2}=
{\cal W}^{\SC (\pm)n \rho}_{j_1 q_1}\bar {\cal W}^{\SC (\pm)n \rho}_{j_2 q_2},
\end{equation}
where ${\cal W}^{\SC (\pm)n \rho}_{j q}$ is the invariant vector
under $U^{\SC (\pm)}$ in the
representation $(n,\rho)$. According to (\ref{wpm}), the magnitude of 
${\cal W}^{\SC (\pm)n \rho}_{j q}$ are given by the volume of $U^{\SC (\pm)}$; as a consequence, 
${\cal W}^{\SC (-) n \rho}_{j q}$ are not normalizable.
\footnote{
The invariant vectors under $U^{\SC (+)}$
are non-vanishing for the representations of the type $(0,\rho)$.   
In this case the invariant vectors can be given explicitly as functions in  
the representation (\ref{iii}), namely
\begin{equation}\nonumber
{\cal W}^{\SC (+) 0 \rho}(z_1,z_2)=(|z_1|^2+|z_2|^2)^{\frac{i}{2}\rho -1}.
\end{equation}
The $U^{\SC (-)}$ case is more subtle, invariant vectors are not zero in the representations
of the type $(0, \rho)$, and $(4k,0)$ ($k=1,2,\dots$)(see \cite{gel}).  
The invariant vectors for the representations $(0,\rho)$
are given by
\begin{equation}\nonumber
 {\cal W}^{\SC (-)0 \rho}(z_1,z_2)=(|z_1|^2-|z_2|^2)^{\frac{i}{2}\rho -1}+
(|z_2|^2-|z_1|^2)^{\frac{i}{2}\rho -1},
\end{equation}
while the ones corresponding to representations of the type $(4k,0)$
($k=1,2,\dots$) are 
\begin{equation}\nonumber
 {\cal W}^{\SC (-)4k 0}(z_1,z_2)=\delta(|z_1|^2-|z_2|^2)\left(\frac{z_1}{\bar z_2}\right)^{2k}.
\end{equation}}
\begin{figure}[h]
\centerline{{\psfig{figure=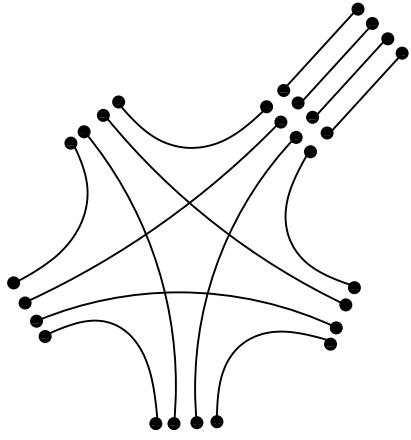,height=6cm}}}
\label{vertex}\bigskip \caption{Structure of the propagator and interaction.  The
black dots represent the projections ${\cal W}^{\SC (\pm)}$ produced by the 
$U^{\SC (\pm)}$ averaging in equation (\ref{mart}).}
\label{figure}
\end{figure}  
Equation (\ref{wpm}) defines a projection operator only with the $+$ sign, 
while in the other case we will have to take care of the infinite volume factor every
time we square (\ref{wpm}). This fact does not cause any problems in our deduction.
This is addressed in section (\ref{app2}). 

One of the two ${\cal W}^{\SC (\pm)n \rho}$ in (\ref{wpm})
appears always contracted with the indices of the $\bar D(\gamma)$
associated to a vertex; while the other is contracted with a
propagator.  We observe that the representation matrices
associated to propagators ($\gamma^{\SC (2)}_{e}$) appear in four faces in
(\ref{aj}).  The ones associated to vertices appear also four times
but combined in the ten corresponding faces converging at a vertex. 
Consequently, they can be paired according to the rule $\bar D^{n \rho}_{j
q k l}(\gamma_{e_i})\bar D^{n \rho}_{k l s t}(\gamma_{e_j}) =\bar D^{n
\rho}(\gamma_{e_i} \gamma_{e_j})_{j q s t}$.  In Fig.\,(\ref{figure})
we represent the structure described above.  A continuous line
represents a representation matrix, while a dark dot a contraction
with an invariant vector ${\cal W}^{\SC (\pm)n \rho}$.  Taking all this into account, and denoting
$\bar D^{n \rho}_{\SC {\cal W}^{\SC (\pm)}{\cal W}^{\SC (\pm)}}(\gamma)$ the matrix element 
$\bar {\cal W}^{\SC (+) n,\rho}_{j_1 q_1}  \bar D^{n,\rho}(\gamma)_{j_1 q_1 j_2 q_2} 
{\cal W}^{\SC (+) n,\rho}_{j_2 q_2}$,
we obtain 
\begin{eqnarray}
    \label{lsf} 
    A^{\SC (\pm)}(J)=\sum_{n_f} \int_{\rho_f} \prod_{{f }}\ (\rho_f^2+n_f^2) \ \
    \prod_{{e }} \ A^{\SC (\pm)}_e( \rho_{\SC e_1},\dots \rho_{\SC e_{4}};n_{\SC
    e_1},\dots n_{\SC e_{4}} ) \ \prod_{{ v }} \ A^{\SC (\pm)}_v(\rho_{\SC v_1},
    \dots \rho_{\SC v_{10}};n_{\SC v_1}, \dots n_{\SC v_{10}}),
    \end{eqnarray} 
where $A^{\SC (\pm)}_e$ is given by
\begin{equation}\label{edge}
A^{\SC (\pm)}_e( \rho_{\SC 1},\dots \rho_{\SC {4}};n_{\SC 1},\dots n_{\SC
{4}})= \int
d\gamma \ \bar D^{n_1 \rho_1}_{\SC {\cal W}^{\SC (\pm)}{\cal W}^{\SC (\pm)}}{\SS (\gamma)} \dots 
\bar D^{n_4 \rho_4}_{\SC {\cal W}^{\SC (\pm)}{\cal W}^{\SC (\pm)}}{\SS (\gamma)},
\end{equation}
and $A^{\SC (\pm)}_{v}$ by
\begin{eqnarray}
\label{choclo}\nonumber && A^{\SC (\pm)}_v(\rho_{\SC
1}, \dots \rho_{\SC {10}};n_{\SC
1}, \dots n_{\SC {10}}) = \\ \nonumber &&
\int \  \prod^{5}_{i=1} d\gamma_i \ 
\bar D^{n_1 \rho_1}_{\SC {\cal W}^{\SC (\pm)}{\cal W}^{\SC (\pm)}}{\SS (\gamma_1\gamma^{-1}_5)} 
\bar D^{n_2 \rho_2}_{\SC {\cal W}^{\SC (\pm)}{\cal W}^{\SC (\pm)}}{\SS (\gamma_1\gamma^{-1}_4)}  
\bar D^{n_3 \rho_3}_{\SC {\cal W}^{\SC (\pm)}{\cal W}^{\SC (\pm)}}{\SS (\gamma_1\gamma^{-1}_3)} 
\bar D^{n_4\rho_4}_{\SC {\cal W}^{\SC (\pm)}{\cal W}^{\SC (\pm)}}{\SS (\gamma_1\gamma^{-1}_2)} \\ &&
\bar D^{n_5 \rho_5}_{\SC {\cal W}^{\SC (\pm)}{\cal W}^{\SC (\pm)}}{\SS (\gamma_2\gamma^{-1}_5)} 
\bar D^{n_6 \rho_6}_{\SC {\cal W}^{\SC (\pm)}{\cal W}^{\SC (\pm)}}{\SS (\gamma_2\gamma^{-1}_4)} 
\bar D^{n_7 \rho_7}_{\SC {\cal W}^{\SC (\pm)}{\cal W}^{\SC (\pm)}}{\SS (\gamma_2\gamma^{-1}_3)} 
\bar D^{n_8 \rho_8}_{\SC {\cal W}^{\SC (\pm)}{\cal W}^{\SC (\pm)}}{\SS (\gamma_3\gamma^{-1}_5)} 
\bar D^{n_9 \rho_9}_{\SC {\cal W}^{\SC (\pm)}{\cal W}^{\SC (\pm)}}{\SS (\gamma_3\gamma^{-1}_4)} 
\bar D^{n_{10} \rho_{10}}_{\SC {\cal W}^{\SC (\pm)}{\cal W}^{\SC (\pm)}}{\SS (\gamma_4\gamma^{-1}_5)} .  
\end{eqnarray} 
In Fig.\,(\ref{figure}), each $\bar D^{n \rho}_{\SC {\cal W}^{\SC (\pm)}{\cal W}^{\SC (\pm)}}(\gamma)$ in
the previous expressions corresponds to a line bounded by two dark dots. Up to this point we have treated
the two theories simultaneously. In the following section we review the properties of the 
$S^{\SC +}[\phi]$-model introduced in \cite{lac}, and then we present the new $S^{\SC +}[\phi]$-model.

\subsection{Lorentzian spinfoam over the Lobachevskian space}

In this section we summarize the results presented in \cite{lac} corresponding to the model
defined by the action $S^{\SC +}[\phi]$ above. 
The functions $\bar D^{n \rho}_{\SC {\cal W}^{\SC (+)}{\cal W}^{\SC (+)}}(\gamma)$ are realized as
functions over the hyperboloid $y^{\mu} y_{\mu}=1$, $y_0>0$ in Minkowski space, known as 
Lobachevskian space (here denoted as $H^{+}$).
They are computed in the following section, using the theory of harmonic  analysis 
over $H^{+}$ defined in \cite{gel}. However, in this case they can be also computed,
in a much more explicit way, using the canonical basis defined in the appendix, and (\ref{proj}) 
(this is a consequence of the definition of the canonical basis which is given
in terms of functions on $SU(2)$, see \cite{lac}). Here we choose the derivation
based on \cite{gel}, because it is easier to extend to the new model in the following
subsection.
According to equation (\ref{pito}) we have 
\begin{equation} 
\bar D^{n \rho}_{\SC {\cal W}^{\SC (+)}{\cal W}^{\SC (+)}}(\gamma_1\gamma^{-1}_2
)= \delta_{n 0} K^{+}_{\rho}(\eta(\gamma_1\gamma^{-1}_2)):=K^{+}_{\rho}(y_1, y_2),
\end{equation}
where $K^{+}_{\rho}(\eta)$ is given in (\ref{popo}).
Finally, the invariant measure on $SL(2,C)$ is simply the product of
the invariant measures of the hyperboloid and $SU(2)$, that is
$d\gamma=du^{+}\, dy$.  Using all this, the vertex and edge amplitudes can
be expressed in simple form.  The edge amplitude (\ref{edge}) becomes 
\begin{equation}
    \label{eee}
A_e(\rho_{\SC 1},\dots \rho_{\SC {4}})=\int dy \
K^{+}_{\rho_1}(y)K^{+}_{\rho_2}(y)K^{+}_{\rho_3}(y)K^{+}_{\rho_4}(y),
\end{equation} where we have dropped the $n$'s from our previous 
notation, since now they all take the value zero.
This expression is finite, and its explicit value is computed in
\cite{BC2}.  Finally, the vertex amplitude (\ref{choclo}) results  
\begin{eqnarray}
    \label{vv}
    \nonumber && A_v(\rho_{\SC 1}, \dots
\rho_{\SC {10}}) = \int dy_1 \dots dy_5 \
K^{+}_{\rho_1}(y_1,y_5)K^{+}_{\rho_2}(y_1,y_4)K^{+}_{\rho_3}
(y_1,y_3)K^{+}_{\rho_4}(y_1,y_2)\\
&& \ \ \ \ \ \ \ \ \ \ \ \ \ \ \ \ \ \ \ \ K^{+}_{\rho_5}(y_2,y_5)
K^{+}_{\rho_6}(y_2,y_4)K^{+}_{\rho_7}(y_2,y_3)
K^{+}_{\rho_8}(y_3,y_5)K^{+}_{\rho_9}(y_3,y_4) K^{+}_{\rho_{10}}(y_4,y_5).
\end{eqnarray}
The previous amplitude is proportional to the infinite volume of the
gauge group $SL(2,C)$. We can now remove this trivial divergence by 
dropping one of the group integrations (see footnote \ref{foot} above).  
The vertex amplitude (\ref{vv}) is precisely the one defined by Barrett 
and Crane in \cite{BC2}.  The spinfoam model is finally given by
\begin{eqnarray}
    A^{\SC (+)}(J)=\int_{\rho_f} d\rho_f \prod_{{f }}\ \rho_f^2 \ \
    \prod_{{e }} \ A^{+}_e( \rho_{\SC e_1},\dots \rho_{\SC e_{4}}) \
    \prod_{{ v }} \ A^{+}_v(\rho_{\SC v_1}, \dots \rho_{\SC v_{10}}),
    \end{eqnarray} 
It corresponds to the Lorentzian model presented in \cite{lac}. As it was argued there,
a more realistic model should include also simple representations of the kind $(n,0)$.
In the following subsection we show that this is the case in the theory defined by
the action $S^{(-)}[\phi]$.

\subsection{Lorentzian spinfoam over the imaginary Lobachevskian space}

Here we introduce the new model defined by the action $S^{\SC (-)}[\phi]$ above.
As it is shown in the next section, the matrix elements 
$\bar D^{n \rho}_{\SC {\cal W}^{\SC (-)}{\cal W}^{\SC (-)}}(\gamma)$ 
in (\ref{edge}) and (\ref{choclo}) are realized as functions on the imaginary Lobachevskian space
(in Minkowski space realized as the 1-sheeted hyperboloid $y^{\mu}y_{\mu}=-1$ where the point $y$ is
identified with $-y$). They correspond to the irreducible components in the harmonic expansion of the delta 
distribution on this space. 
According to equation (\ref{++})   
\begin{equation}
\bar D^{n \rho}_{\SC {\cal W}^{\SC (-)}{\cal W}^{\SC (-)}}(\gamma_1\gamma^{-1}_2
)=K^{-}_{n \rho}(\eta(\gamma_1\gamma^{-1}_2),\hat r_z(\gamma_1\gamma^{-1}_2)):=K^{-}_{n \rho}(y_1, y_2), 
\end{equation}
where $K^{-}_{n \rho}(\eta, \hat r_z)$ is defined as
\begin{equation}
K^{-}_{n \rho}(\eta, \hat r_z)=\delta_{n 0} K^{-}_{\rho}(\eta, \hat r_z)+\delta_{4 k 0}\delta(\rho)K^{-}_{4 k}(\eta, \hat r_z),
\end{equation} according to (\ref{carro}) and (\ref{carron}).
The invariant measure on $SL(2,C)$ is simply the product of
the invariant measures of the imaginary Lobachevskian space and $U^{\SC (-)}$ respectively, that is
$d\gamma=du^{-}\, dy$. The edge and vertex amplitudes then become 
\begin{equation}
    \label{eee-}A_e(\rho_{\SC 1},\dots \rho_{\SC {4}};n_{\SC 1},\dots n_{\SC {4}} )=\int dy \
K^{-}_{n_1 \rho_1}(y) K^{-}_{n_2 \rho_2}(y)K^{-}_{n_3\rho_3}(y)K^{-}_{n_4 \rho_4}(y),
\end{equation} and
\begin{eqnarray}
    \label{vv-}
    \nonumber && A^{\SC -}_v(\rho_{\SC 1}, \dots
\rho_{\SC {10}};n_{\SC 1}, \dots
n_{\SC {10}}) = \int dy_2 \dots dy_5 \
K^{-}_{n_1 \rho_1}(y_1,y_5)K^{-}_{n_2 \rho_2}(y_1,y_4)K^{-}_{n_3\rho_3}
(y_1,y_3)K^{-}_{n^4 \rho_4}(y_1,y_2)\\
&& \ \ \ \ \ \ \ \ \ \ \ \ \ \ \ \ \ \ \ \ K^{-}_{n_5 \rho_5}(y_2,y_5)
K^{-}_{n_6 \rho_6}(y_2,y_4)K^{-}_{n_7 \rho_7}(y_2,y_3)
K^{-}_{n_8 \rho_8}(y_3,y_5)K^{-}_{n_9 \rho_9}(y_3,y_4) K^{-}_{n_10 \rho_{10}}(y_4,y_5)
\end{eqnarray}respectively.
Where as in the previous model we have dropped one of the integration variables from the definition 
of the vertex amplitude to remove a trivial infinite volume factor. In the spinfoam model defined 
by $S^{\SC (-)}[\phi]$ the amplitude of a given 2-complex $J$ is finally
\begin{eqnarray}\label{momodel}
    A^{\SC (-)}(J)=\sum_{n_f}\int_{\rho_f} d\rho_f \prod_{{f }}\ (\rho_f^2 + n^2_f)  \
    \prod_{{e }} \ A^{-}_e( \rho_{\SC e_1},\dots \rho_{\SC e_{4}};n_{\SC e_1},\dots n_{\SC e_{4}}) \
    \prod_{{ v }} \ A^{-}_v(\rho_{\SC v_1}, \dots \rho_{\SC v_{10}};n_{\SC v_1}, \dots n_{\SC v_{10}}).
\end{eqnarray} 

\section{Calculation of the kernels $K^{\SC\pm}$}\label{app2}

In this section we compute the values of the projector $K^{\SC\pm}$
which appear in the vertex and edge amplitudes of the two models. 

\subsection{Functions on the Lobachevskian space}

Take the 2-sheeted hyperboloid on Minkowski space 
defined by $x^{\nu}x_{\nu}=1$. Every point on upper sheet of the hyperboloid 
can be written as \begin{equation}\label{iso}
x_g = g g^{\dagger}.
\end{equation}
This space is a possible realization of the Lobachevskian space (from now on denoted by $H^{+}$).
The theory of harmonic analysis on this space is relevant for
the spinfoam model defined by the action $S^{\SC (+)}[\phi]$ as can be seen 
from the fact that $H^{+}=SL(2,C)/SU(2)$. 
Given a square integrable function $\tilde f(g)$ over $SL(2,C)$ we can define a 
square integrable function $f(x) \in {\cal L}^2 (H^{\SC +})$ by means of 
averaging $\tilde f$ over the subgroup $U^{\SC (+)}=SU(2)$.
Namely,
\begin{equation}\label{mar}
f(x)=\int_{U^{\SC(+)}}\tilde f(g_x u^{\SC (+)}) du^{\SC (+)},
\end{equation}
where $g_x \in SL(2,C)$ represents the equivalence class
of transformations  taking the apex $(1,0,0,0)$ to the point
$x$ on the hyperboloid. Expanding $\tilde f(g)$ in modes,
using (\ref{pass}), we can construct the theory of 
harmonic analysis of $f(x)$. Thus
\begin{equation}
f(x)=\sum^{\infty}_{n=0} \int^{\infty}_{\rho=0} \int_{U^{\SC (+)}}
\bar  D^{n,\rho}_{j_1 q_1 j_2 q_2}(g_x u^{\SC (+)}) \tilde f^{j_1 q_1
j_2 q_2}_{n,\rho}(n^2+\rho^2) d\rho d u^{\SC (+)}, 
\end{equation}
then using (\ref{wpm}) and defining $f^{j_1 q_1}_{n,\rho}:=
\tilde f^{j_1 q_1 j q}_{n,\rho} \bar {\cal W}^{\SC (+) n,\rho}_{j q}$ we obtain
\begin{equation}\label{vito}
f(x)=\sum^{\infty}_{n=0} \int^{\infty}_{\rho=0}
\left[ \bar  D^{n,\rho}_{j_1 q_1 j_2 q_2} (g_x) {\cal W}^{\SC (+) n,\rho}_{j_2 q_2} \right]  
f^{j_1 q_1}_{n,\rho} (n^2+\rho^2) d\rho.
\end{equation}
Finally using (\ref{furie}),(\ref{wpm}), and the fact that any $g\in SL(2,C)$ can be written 
as $g_x u^{\SC (+)}$ 
\begin{eqnarray}\label{tambien}
\nonumber f^{j_1 q_1}_{n,\rho}&=&\int \tilde f(g_x u^{\SC (+)}) D^{n,\rho}_{j_1 q_1 j q}(g_x u^{\SC (+)})
\bar {\cal W}^{\SC (+) n,\rho}_{j_2 q_2} d(g_xu^{\SC (+)})\\  \nonumber
&=& \int \tilde f(g_x u^{\SC (+)}) D^{n,\rho}_{j_1 q_1 j q}(g_x) 
\bar {\cal W}^{\SC (+) n,\rho}_{j_2 q_2} dx du^{\SC (+)} \\
&=&\int f(x) \left[ D^{n,\rho}_{j_1 q_1 j q}(g_x)\bar {\cal W}^{\SC (+) n,\rho}_{j q}\right] dx.
\end{eqnarray}
In the second line we used $d(g_xu^{\SC (+)})=dxdu^{\SC (+)}$, where $dx$ 
is the invariant measure on $H^{\SC (+)}$ and invariance of ${\cal W}^{\SC (+) n,\rho}$ under $U^{\SC +}$.
In the last line we used equation (\ref{mar}).
Combining these two equations we can write 
\begin{eqnarray}\label{deltita}
\delta(x)=\sum^{\infty}_{n=0} \int^{\infty}_{0} (n^2+ \rho^2) d\rho \; 
\ \bar {\cal W}^{\SC (+) n,\rho}_{j_1 q_1} \bar D^{n,\rho}(g_x)_{j_1 q_1 j_2 q_2} 
{\cal W}^{\SC (+) n,\rho}_{j_2 q_2}. 
\end{eqnarray}
The term in brackets in (\ref{vito}) and (\ref{tambien}) corresponds to the 
analogous of spherical harmonics in the case of functions over $S^2=SU(2)/U(1)$. 
They are different from zero only for simple irreducible representations of the type $(0,\rho)$.
This can be explicitly shown using the properties of the canonical basis 
defined in the appendix (see also \cite{lac}). However, this method can not 
be easily extended to the case which is relevant to the new model introduced in the paper.  
We therefore complete the construction in the more general framework of 
reference \cite{gel}, where the theory of harmonic analysis over the $SL(2,C)$ 
homogeneous spaces is defined with the aid of integral geometry elements.
 
Given $f(x) \in {\cal L}^2 (H^{\SC +})$, square integrable function over the 
Lobachevskian space, its Gelfand transform is defined as
\begin{equation}\label{fufu}
F(\xi;\rho)=\int_{H^{-}} f(x) \left(x^{\nu}\xi_{\nu}\right)^{i\rho/2-1} dx,
\end{equation} 
where $\rho \in [0,\infty)$ and $\xi$ is a vector on the null cone $C^{\SC +}$
normalized such that $\xi_0=1$. The invariant measure $dx$ on the hyperboloid is defined 
up to a constant factor that we choose in order to simplify some equations (our measure 
differs from the one in \cite{gel} by a $(4\pi)^2$ factor). 
It turn out that the function $F(\xi;\rho)$ lives in  the irreducible
representation of the type $(0,\rho)$ of $SL(2,C)$. Moreover, it corresponds to the Fourier 
component of $f(x)$ in the balance representation $(0,\rho)$ as can be explicitly
seen writing the inversion formula for (\ref{fufu}). Choosing coordinates this inversion formula can be written as 
\begin{eqnarray}\label{fhy+}
f(x)= \int^{\infty}_0 \rho^2 d\rho \int_{C^{\SC +}} F(\xi;\rho)
\left(x^{\nu} \xi_{\nu} \right)^{-i\rho/2-1} d\omega, 
\end{eqnarray}  
where 
\begin{equation}
d\omega=\frac {1}{4\pi}{\rm sin}(\theta) d\theta d\phi,
\end{equation}
is the normalized measure on the sphere defined by the null cone with normalization $\xi_0=1$.  
Combining (\ref{fhy+}) with (\ref{fufu}) the resolution of the identity on $H^{+}$ 
becomes
\begin{eqnarray}\label{mmm}
\delta(x,y)=  \int^{\infty}_0 \rho^2 d\rho \int_{\Gamma^{+}}
\left(y^{\nu} \xi_{\nu} \right)^{i\rho/2-1}  
\left(x^{\nu} \xi_{\nu} \right)^{-i\rho/2-1} d\omega 
\end{eqnarray}  The previous equation can be express as a sum over the projectors
$K^{\SC (+)}_{\rho}(x,y)$ over the representation $(0,\rho)$, namely
\begin{equation}\label{mmmm}
\delta(x,y)=\int^{\infty}_0 \rho^2 d\rho K_{\rho}(x,y),
\end{equation}
where
\begin{equation}
K^{\SC (+)}_{\rho}(x,y)=  \int_{\Gamma^{+}}
\left(y^{\nu} \xi_{\nu} \right)^{i\rho/2-1}  
\left(x^{\nu} \xi_{\nu} \right)^{-i\rho/2-1} d\omega.  
\end{equation}
It is easy to see that the previous function depends only on the 
hyperbolic distance between $x$ and $y$. To compute explicitly its value 
we chose $x$ to be the hyperboloid apex ($x=(1,0,0,0)$), while we take 
$y=({\rm cosh}(\eta),0,0,{\rm sinh}(\eta))$. In our normalization the
4-vector $\xi$ can be written as
\begin{equation}
\xi=\left(1,{\rm sin}(\theta){\rm cos}(\phi), {\rm sin}(\theta){\rm sin}(\phi), {\rm cos}(\theta) \right).
\end{equation}
With all this the previous equation becomes
\begin{eqnarray}\label{popo}
\nonumber K^{\SC (+)}_{\rho}(\eta)&=& \frac{1}{4\pi} \int \left({\rm cosh}(\eta)-{\rm cos}
(\theta){\rm sinh}(\eta) \right)^{i\rho/2-1}
{\rm sin}(\theta) d\theta d\phi \\ 
&=& {2 \; {\rm sin}({1 \over 2} \rho  \eta) \over \rho \; {\rm sinh}(\eta)}
\end{eqnarray}
Comparing equation (\ref{deltita}) with (\ref{mmmm}) we conclude that
\begin{equation}\label{pito}
D^{n,\rho}_{\SC {\cal W}^{\SC (+)}{\cal W}^{\SC (+)}}(g_y):=\bar {\cal W}^{\SC (+) n,\rho}_{j_1 q_1} \bar D^{n,\rho}(g_y)_{j_1 q_1 j_2 q_2} {\cal W}^{\SC (+) n,\rho}_{j_2 q_2}= \delta_{n 0} K^{\SC (+)}_{\rho}(x,y) 
\end{equation}

\subsection{Functions on the imaginary Lobachevskian space\label{app1}}

The 1-sheeted hyperboloid on Minkowski space is given by the points 
such that $x^{\nu}x_{\nu}=-1$. Every point on the hyperboloid 
can be written as 
\begin{equation}\label{isop}
x_g = g \sigma_3 g^{\dagger}.
\end{equation}
The imaginary Lobachevskian space $H^{\SC -}$ corresponds to the 1-sheeted hyperboloid 
where the point $x$ is identified with $-x$. This is exactly the homogeneous space 
$SL(2,C)/U^{\SC (-)}$ relevant for the construction of the new model presented in the paper. 
In analogy to the previous section we can defined a function on 
the imaginary Lobachevskian space by means of averaging functions on $SL(2,C)$ over 
the sub-group $U^{\SC (-)}=SU(1,1)\times Z_2$. Given $\tilde f(g) \in {\cal L}^2(SL(2,C))$
we can define $f(x)\in {\cal L}^2(H^{\SC (-)})$ as 
\begin{equation}\label{tincho}
f(x)=\int_{U^{\SC(-)}}\tilde f(g_x u^{\SC (-)}) du^{\SC (-)},
\end{equation}
where $g_x \in SL(2,C)$ represents the equivalence class
of transformations  taking the point $(0,0,0,1)$ to the point
$x$ on the hyperboloid. The difficulty now is that the sub-group $U^{\SC (-)}$ is no longer compact. 
As a consequence the RHS of the previous equation is not a square 
integrable as a function on $SL(2,C)$.
Expanding $\tilde f(g)$ in modes, using (\ref{pass}), 
\begin{equation}
f(x)=\sum^{\infty}_{n=0} \int^{\infty}_{\rho=0} \int_{U^{\SC (-)}}
\bar  D^{n,\rho}_{j_1 q_1 j_2 q_2}(g_x u^{\SC (-)}) \tilde f^{j_1 q_1
j_2 q_2}_{n,\rho}(n^2+\rho^2) d\rho d u^{\SC (-)}, 
\end{equation}
then using (\ref{wpm}) and defining $f^{j_1 q_1}_{n,\rho}:=
\tilde f^{j_1 q_1 j q}_{n,\rho} {\cal W}^{\SC (-) n,\rho}_{j q}$ we obtain
\begin{equation}\label{vitos}
f(x)=\sum^{\infty}_{n=0} \int^{\infty}_{\rho=0}
\left[ \bar  D^{n,\rho}_{j_1 q_1 j_2 q_2} (g_x) {\cal W}^{\SC (-) n,\rho}_{j_2 q_2} \right]  
f^{j_1 q_1}_{n,\rho} (n^2+\rho^2) d\rho, 
\end{equation}
where 
\begin{eqnarray}\label{tambie}
\nonumber f^{j_1 q_1}_{n,\rho}&=&\int \tilde f(g_x u^{\SC (-)}) D^{n,\rho}_{j_1 q_1 j q}(g_x u^{\SC (-)})
\bar {\cal W}^{\SC (-) n,\rho}_{j_2 q_2} d(g_xu^{\SC (-)})\\  \nonumber
&=& \int \tilde f(g_x u^{\SC (-)}) D^{n,\rho}_{j_1 q_1 j q}(g_x) 
\bar {\cal W}^{\SC (-) n,\rho}_{j_2 q_2} dx du^{\SC (-)} \\
&=&\int f(x) \left[ D^{n,\rho}_{j_1 q_1 j q}(g_x)\bar {\cal W}^{\SC (-) n,\rho}_{j q}\right] dx.
\end{eqnarray}
The non-compactness of $U^{\SC (-)}$ implies the non-normalizability of the ${\cal W}^{\SC (-) n,\rho}$ 
invariant vectors. Their presence in (\ref{pass}) correspond to a distributional factor in the 
Fourier components of $f(x)$ thought as a $SL(2,C)$ function.
\footnote{A simple analogy corresponds to the following example in
Fourier analysis on $\Re^2$.  Take the square integrable function
$\tilde f(x,y)={\rm exp}(-x^2-y^2)$.  The analogous to the group
$SL(2,C)$ is here the group of translations in $\Re^2$.  Every
function $\tilde f(x,y)$ can be thought of as a function of the group
element that takes the origin into the point $(x,y)$.  We can define
an invariant function under the action of the subgroup $U$ of
translations in the $y$ direction by averaging $\tilde f(x,y)$ under
the action of $U$, namely
\begin{eqnarray}
\nonumber f(x)=\int_{-\infty}^{\infty} \tilde f(x,y+u) du = {\sqrt \pi} e^{-x^2},
\end{eqnarray}
where in the last equality we have used our Gaussian example for
$\tilde f(x,y)$.  The function $f(x)$ is a perfectly square integrable
function of one variable but is no longer square integrable on
$\Re^2$.  Writing the previous equation in momentum space we have
\begin{eqnarray}
\nonumber f(x)&=&\frac{1}{4\pi^2}\int \tilde F(k_x,k_y) e^{i(x k_x +
(y+u) k_y)}du dk_x dk_y \\
\nonumber &=& \frac{1}{2\pi}\int \tilde F(k_x,0) e^{i x k_x } dk_x ,
\end{eqnarray} 
where 
\begin{eqnarray}
\nonumber
\tilde F(k_x,0)=\int \tilde f(x,y) e^{-i x k_x} dx dy =\int f(x) e^{-i x k_x} dx
\end{eqnarray}
is the Fourier component of $f(x)$ in one dimension.
If we think of $f(x)$ as a function on $\Re^2$ then its Fourier transform takes 
distributional values in $k_y$, namely $F(k_x,0) \delta(k_y)$.}
Combining the last two equations we can read off the expression for the delta
distribution on $H^{\SC (-)}$, namely
\begin{eqnarray}\label{deltota}
\delta(x)=\sum^{\infty}_{n=0} \int^{\infty}_{0} (n^2+ \rho^2) d\rho \; 
\ \bar {\cal W}^{\SC (-) n,\rho}_{j_1 q_1} \bar D^{n,\rho}(g_x)_{j_1 q_1 j_2 q_2} 
{\cal W}^{\SC (-) n,\rho}_{j_2 q_2}. 
\end{eqnarray}

Now we present the results of \cite{gel} on the harmonic analysis on $H^{\SC -}$
based on Gelfand's integral transforms. Given $f(x)\in {\cal L}^2(H^{\SC -})$, 
square integrable function over the 
imaginary Lobachevskian space, it can be expanded 
in terms of its irreducible $SL(2,C)$ components as
\begin{eqnarray}\label{fhy}
\nonumber && f(x)=  \int^{\infty}_0 \rho^2 d\rho \int_{\Gamma^{+}} F(\xi;\rho)
\left|x^{\nu} \xi_{\nu} \right|^{-i\rho/2-1} d\omega + \\
 && \ \ \ \ \ \ \ \ \ \ \ \ \ \ \ \ \ \ \ \ \ \ \ \ \ \ 32\, \pi
 \sum^{\infty}_{k=1} (4 k)^2 \int_{C^{+}} F(\xi,x;2k) \delta(x^{\nu}
 \xi_{\nu}) d\omega,
\end{eqnarray}  
where $C^{+}$, and $d\omega$ are defined as in the previous section.
The functions  $F(\xi;\rho)$ and $ F(\xi,x;2k)$ correspond to the Fourier 
components in the balance representation $(0,\rho)$ and $(4k,0)$ respectively. 
They are explicitly given by
\begin{equation}
F(\xi;\rho)=\int_{H^{-}} f(x) \left|x^{\nu}\xi_{\nu}\right|^{i\rho/2-1} dx,
\end{equation} 
and
\begin{equation}
F(\xi,x; 2k)=\frac{1}{k}\int_{H^{-}} f(y) e^{-2 i k \Theta(x,y)}\delta(y^{\nu}\xi_{\nu}) dy,
\end{equation}
where the function $\Theta(x,y)$ is  defined by the following equation
\begin{equation}\label{TT}
{\rm cos}(\Theta)=\left|x^{\nu}y_{\nu} \right|.
\end{equation}   
The geometric interpretation of the angle $\Theta$ defined in the
previous equation can be done as follows.  The line generators of the
1-sheeted hyperboloid correspond to null geodesics known as isotropic
lines\cite{gel}.  For each point $x\in H^{-}$ the null geodesic
$x(\lambda)=x+\lambda \xi$ for $\lambda \in \Re$ and $\xi \in C^{+}$
such that $x^{\mu}\xi_{\mu}=0$ is on $H^{-1}$ for all values of
$\lambda$.  Fixing $x$ there is a circle worth of such lines. 
Consider a second point $y$ and search for an isotropic line
containing $y$ and parallel to one crossing $x$ (i.e., having the same
null generator $\xi$).  In order for it for exist, we need
$x^{\mu}\xi_{\mu}=y^{\mu}\xi_{\mu}=0$.  In general, the two lines will
intersect the sphere given by the section $x_0=0$ of the hyperboloid
in two different points.  One can easily verify that the scalar
product $x(\lambda_1)^{\mu}y_{\mu}(\lambda_2)$ is independent of the
values of $\lambda_1$ and $\lambda_2$.  Therefore we can calculate the
scalar product of eq.(\ref{TT}) at the $\lambda$-values for which the
two lines intersect the sphere.  By doing that we conclude that the
value of $\Theta(x,y)$ corresponds to the azimuthal separation of
those points on the sphere.  Combining (\ref{fhy}) with previous
equations the resolution of the identity on $H^{-}$ becomes
\begin{eqnarray}\label{pipo}
&& \nonumber \delta(x,y)=  \int^{\infty}_0 \rho^2 d\rho \int_{\Gamma^{+}}
\left|y^{\nu} \xi_{\nu} \right|^{i\rho/2-1}  
\left|x^{\nu} \xi_{\nu} \right|^{-i\rho/2-1} d\omega + \\
 && \ \ \ \ \ \ \ \ \ \ \ \ \ \ \ \ \ \ \ \ \ \ \ \ \ \ 32\, \pi
 \sum^{\infty}_{k=1} (4k)^2 \int_{\Gamma^{+}} e^{-2i k
 \left[\Theta(x,y) \right]} \delta(x^{\nu} \xi_{\nu})\delta(y^{\nu}
 \xi_{\nu}) d\omega.
\end{eqnarray}  
The previous equation can be express as a ``sum'' over the projectors
$K^{\SC (-)}_{\rho}(x,y)$ and $K^{\SC (-)}_{4 k}(x,y)$, namely
\begin{equation}\label{ppp}
\delta(x,y)=\int^{\infty}_0 \rho^2 d\rho K^{\SC (-)}_{\rho}(x,y)
+\sum^{\infty}_{k=1} (4 k)^2 K^{\SC (-)}_{4 k}(x,y),
\end{equation}
where
\begin{equation}\label{carro}
K^{\SC (-)}_{\rho}(x,y)=  \int_{\Gamma^{+}}
\left|y^{\nu} \xi_{\nu} \right|^{i\rho/2-1}  
\left|x^{\nu} \xi_{\nu} \right|^{-i\rho/2-1} d\omega,  
\end{equation}
and
\begin{equation}
K^{\SC (-)}_{4 k}(x,y)=\frac{32\, \pi}{k}
\int_{\Gamma^{+}}  e^{-2i k \left[\Theta(x,y) \right]} 
\ \delta(x^{\nu} \xi_{\nu})\ \delta(y^{\nu} \xi_{\nu}) d\omega.
\end{equation}
Lets analyze equation (\ref{carro}); we take $\xi=\left(1,{\rm
sin}(\theta){\rm cos}(\phi), {\rm sin}(\theta){\rm sin}(\phi), {\rm
cos}(\theta) \right)$, and $x=(0,0,0,1)$.  We parameterize $y$ in the
following way
\begin{equation}
y=\left({\rm sinh}(\eta),{\rm cosh}(\eta)\ \hat r \right),
\end{equation}      
where $\hat r$ represents a point on the unit sphere. In terms of these coordinates 
(\ref{carro}) becomes
\begin{eqnarray}
&&\nonumber K^{\SC (-)}_{\rho}(\eta,\hat r)=\frac{1}{4\pi} \int {\rm
sin}(\theta) \, d\theta d\phi \times\\
&&\nonumber \left|{\rm sinh}(\eta)- {\rm cosh}(\eta) \left[{\rm
sin}(\theta){\rm cos}(\phi)\hat r_x + {\rm sin}(\theta){\rm
sin}(\phi)\hat r_y + {\rm cos}(\theta) \hat r_z \right]
\right|^{i\rho/2-1} \left| {\rm cos}(\theta) \right|^{-i\rho/2-1} =\\
&&  \int \, dt d\phi
\left|{\rm sinh}(\eta)- {\rm cosh}(\eta)
\left[(1-t^2)^{(1/2)}\left({\rm cos}(\phi)\hat r_x + 
{\rm sin}(\phi)\hat r_y \right) + t \hat r_z   \right]  \right|^{i\rho/2-1}  
\left| t \right|^{-i\rho/2-1}.
\end{eqnarray}
Finally using that $A\; {\rm sin}(\alpha )+ B\; {\rm cos}(\alpha )=
(A^2+B^2)^{1/2} {\rm sin}(\alpha+\alpha_0)$ we obtain \begin{eqnarray}
K^{\SC (-)}_{\rho}(\eta,\hat r)= \frac{1}{4\pi} \int \, dt d\phi
\left|{\rm sinh}(\eta)- {\rm cosh}(\eta) \left[(1-t^2)^{(1/2)}(1-\hat
r_z^2)^{(1/2)} {\rm sin}(\phi) + t \hat r_z \right]
\right|^{i\rho/2-1} \left| t \right|^{-i\rho/2-1} .
\end{eqnarray}
From the previous equation we conclude that $K^{\SC
(-)}_{\rho}(\eta,\hat r)$ behaves asymptotically as
\begin{equation}
K^{\SC (-)}_{\rho}(\eta,\hat r)=\tilde \alpha_{\rho} (\hat r_z)\; e^{-
\eta}\, e^{i\eta \rho/2},
\end{equation}
where $\tilde \alpha(\hat r_z)$ is an integral of a finite function
over the compact space $[0,1]\times S^1$ and therefore is finite. 
When $\hat r_z=1$ the integral above can be performed and we obtain
\begin{equation}\label{carrito}
K^{\SC (-)}_{\rho}(\eta,\hat r_z=1)= {2{\rm sin}({1 \over 2} \rho
\eta) \over \rho \; {\rm sinh}(\eta)}.
\end{equation}
When the points $x$ and $y$ lay on the same boost orbit the projector
$K^{\SC (-)}_{\rho}(x,y)$ has the same form as $K^{\SC
(+)}_{\rho}(x,y)$.  In the same parametrization the projector $K^{\SC
(-)}_{4 k}(x,y)$ becomes
\begin{eqnarray}
\nonumber K^{\SC (-)}_{4 k}(\eta,\hat r_z) &=& \frac{8\, e^{-2i k
\Theta(\eta,\hat r_z)}}{k}\int \delta\left({\rm sinh}(\eta)- {\rm
cosh}(\eta) (1-\hat r_z^2)^{(1/2)} {\rm sin}(\phi) \right) d\phi \\
&=& {8\, e^{-2ik\Theta(\eta,\hat r_z)}\over k \left( 1-\hat r^2_z {\rm
cosh}^2(\eta)\right)^{1/2}} \int \delta\left( \phi-{\rm arcsin}
\left[{{\rm tanh}(\eta) \over (1-\hat r_z^2)^{(1/2)}}\right] \right)
d\phi.  \end{eqnarray} The integral on the right vanishes unless
$\left|{{\rm tanh}(\eta) \over (1-\hat r_z^2)^{(1/2)}}\right| \le 1$. 
Notice that therefore $K^{\SC (-)}_{4 k}(\eta,\hat r_z)$ has support
on the values of $\eta$, and $\hat r_z$ for which $\Theta={\rm
arcos}(|{\rm cosh}(\eta)\hat r_z|)$ is real with a range $0\le \Theta
\le \frac{\pi}{2}$.  Using that ${\rm cos}(\Theta)=|{\rm cosh}(\eta)
\hat r_z|$ (equation (\ref{TT})) we finally obtain
\begin{equation}\label{carron}
K^{\SC (-)}_{4 k}(\eta,\hat r_z)={8\, e^{-2ik \Theta}\over k \; {\rm
sin}(\Theta)}\end{equation} for $0\le \Theta(\eta,\hat r_z) \le \frac
{\pi}{2}$ and zero otherwise.  The real part of the previous projector
diverges at $\Theta=0$, i.e., on the curve $\hat r_z = \pm {\rm
cosh}^{-1}(\eta)$.  If we define $J=k-\frac {1}{2}$ then $J$ takes all
the half-integer values and the imaginary part of $K^{\SC (-)}_{4
k}(\eta,\hat r_z)$ becomes
\begin{equation}\label{ima}
{\rm Im}\left[K^{\SC (-)}_{4 k}(\eta,\hat r_z)\right]=\frac {8}{J+\frac {1}{2}}\ \frac {{\rm sin}\left((2J+1)\Theta\right)}
{{\rm sin}(\Theta)}
\end{equation} for $0\le \Theta \le \pi$ and vanishes otherwise. The imaginary part 
of the projector on the discrete representations has the same form as the one appearing
in the Euclidean Barrett-Crane models.   
Finally, comparing (\ref{deltota}) with (\ref{ppp}) for $x=(0,0,0,1)$ we conclude that
\begin{equation}\label{++}
D^{n,\rho}_{\SC {\cal W}^{\SC (-)}{\cal W}^{\SC (-)}}(g_y)=
\delta_{n 0} K^{\SC (-)}_{\rho}(x,y) + \delta_{n, 4 k} \delta(\rho) 
K^{\SC (-)}_{4 k}(x,y).
\end{equation}

\section{Discussion}

We have carried over a generalization of the model defined in
\cite{lac}.  The new model is given by an $SL(2,C)$ BF quantum theory
plus a quantum implementation of the constraints that reduce BF theory
to Lorentzian general relativity.  This corresponds to the restriction
to simple representations, those for which the Casimir $\hat {\cal
C}_2$ vanishes (see (\ref{c2})).  In contrast with the previous model,
the present one includes also elements of the discrete series in the
set of simple representations. 

Four dimensional Lorentzian quantum spacetime appears as a fully
combinatorial notion represented by spinfoams colored by simple
representations of $SL(2,C)$.  The model possesses an intrinsically
defined local causal structure with is non-perturbative and background
independent.  Causality in the model is induced by the algebra of
$SL(2,C)$ (see equations (\ref{c1}) and (\ref{c2})).  In particular,
the Casimir (\ref{c1}) can be interpreted as the square of the area
operator corresponding to quantized bivectors \cite{Baez}.  Space-like
and time-like bivectors can be classified according to the two
possibilities $\hat {\cal C}_1 > 0$ or $\hat {\cal C}_1 < 0$.  We
define a space-like section of a given spinfoam as the colored graph
(spin network) defined by the intersection of a 3-surface with the
corresponding 2-complex such that it is labeled by simple
representations in the discrete series.  On this representations the
area operator $\hat {\cal A}$ reduces to
\begin{equation}
\hat{\cal A} \  \sim \  {\sqrt {J(J+1)}}\  \hat 1,
\end{equation} 
where $J:=k-1/2$ $(k=1,2 \dots)$ and therefore takes only half-integer
values.  Notice that the spectrum of the area on spatial sections of
the spinfoams of the model is contained in the one predicted by Loop
Quantum Gravity in the canonical formalism\cite{area}.
\footnote{Our definition of space-like section should be in agreement
with the geometrical analysis of the discretization of BF theory with
the corresponding passage to GR through the implementation of
Plebanski's constraints.  As it is pointed out in \cite{BC2} this is a
delicate issue.  To answer this question a rigorous definition of the
area operator as well as a deeper understanding of the geometry of the
quantization prescription is needed.  This important issue will be
study in the future.}
Only those eigenvalues corresponding to half integer spin appear. 
This is associated to the fact that the model was based on the
harmonic analysis of even functions on the one sheeted hyperboloid,
the imaginary Lobachevkian space.  The extension of the model
including the other part of the spectrum seems possible and it will be
study elsewhere.

A necessary condition for the model to be well defined is the
finiteness of the edge and vertex amplitudes appearing in
(\ref{momodel}).  Both amplitudes turn out to be finite in the $S^{\SC
+}[\phi]$ model \cite{BB}.  We do not address this issue in the paper,
but we want to make a few comments on some of the results that might
be relevant for future research in this respect.  In the new model the
projectors $K^{\SC (-)}_{\rho}(x,y)$ have the same asymptotic behavior
of $K^{\SC (+)}_{\rho}(x,y)$ (they even coincide when the two points
are located in the same spatial direction (see eq.  (\ref{carrito})). 
Therefore, one would expect no divergences coming from this kind of
projectors.  The analysis of the discrete projectors $K^{\SC (-)}_{4
k}(x,y)$ is more delicate.  The imaginary part
\begin{equation}
{\rm Im}\left[K^{\SC (-)}_{4 k}(x,y)\right] \sim \frac {{\rm
sin}\left((2J+1)\Theta(x,y)\right)} {{\rm sin}(\Theta(x,y))}
\end{equation}
is well behaved and has the same functional form of the corresponding
projector in the Euclidean Barrett-Crane model\cite{barret}.  The real
part diverges on $\Theta=0$ which according to the geometrical
interpretation given in the last section corresponds to the situation in
which $x$ and $y$ lay on the same null generator of the hyperboloid. 
The study of the convergence of the different amplitudes appearing in
the new model is left for future studies.
 
Another problem is the the convergence of the sum over representations
in (\ref{momodel}).  This problem appears also in the Euclidean
models.  To cure it, the Barrett-Crane model was defined in terms of a
quantum deformation of the gauge group ($SO_q(4)$, with $q^n =1$). 
The quantum deformation introduces a cut-off in the summ over
representations that regularizes the amplitudes.  In the limit in
which the quantum deformation is removed ($q \rightarrow 1$),
divergences appear whenever the 2-complex $J$ includes bubbles.  A
similar regularization for the Lorentzian state sum model is suggested
by Barrett and Crane in \cite{BC2}.  A different strategy for dealing
with this infinity was suggested in reference \cite{ac}, using the
field theory over group technology.  In this reference, we have
defined a natural variant of the Euclidean Barrett-Crane model based
on a different implementation of the BF-to-GR constraints which,
however, turns out to be finite \cite{fac}.  On this model, see also
\cite{chor}.  The Lorentzian model presented in \cite{lac} as well
as its extension presented here correspond to the finite version of
the Euclidean model.  Accordingly, although further study is certainly
needed, we suspect that the Lorentzian models presented here might
also be finite.

\section{Acknowledgments}  

A.P. wants to specially thank Ted Newman for his illuminating introduction
to the representation theory of the Lorentz group and much more. 
This work was partially supported by NSF Grant PHY-9900791. 

\begin{appendix}
      
\section{Representation Theory of $SL(2,C)$}

We review a series of relevant facts about $SL(2,C)$ representation 
theory. Most of the material of this section can be found in \cite{gel,ru}.
For a very nice presentation of the subject see also \cite{ted}.

We denote an element of $SL(2,C)$ by
\begin{equation}\label{g}
g=\left[\begin{array}{c} \alpha \ \beta \\
\gamma \ \delta  \end{array}\right],
\end{equation}
with $\alpha$, $\beta$, $\gamma$, $\delta$ complex numbers such that
$\alpha \delta-\beta \gamma=1$.  All the finite dimensional
irreducible representations of $SL(2,C)$ can be cast as a
representation over the set of polynomials of two complex variables
$z_1$ and $z_2$, of order $n_1-1$ in $z_1$ and $z_2$ and of order
$n_2-1$ in $\bar z_1$ and $\bar z_2$.  The representation is given by
the following action
\begin{equation}
T(g)P(z_1,z_2)=P(\alpha z_1 +\gamma z_2,\beta z_1 +\delta z_2).
\end{equation}
The usual spinor representations can be directly related to these ones.
 
The infinite dimensional representations are realized over the space
of homogeneous functions of two complex variables $z_1$ and $z_2$ in
the following way.  A function $f(z_1, z_2)$ is called homogeneous of
degree $(a,b)$, where $a$ and $b$ are complex numbers differing by an
integer, if for every $\lambda \in C$ we have
\begin{equation}\label{hom} f(\lambda z_1, \lambda z_2)=\lambda^a
\bar\lambda^b f(z_1, z_2),
\end{equation}
where $a$ and $b$ are required to differ by an integer in order to
$\lambda^a \bar\lambda^b$ be a singled valued function of $\lambda$. 
The infinite dimensional representations of $SL(2,C)$ are given by the
infinitely differentiable functions $f(z_1, z_2)$ (in $z_1$ and $z_2$
and their complex conjugates) homogeneous of degree $({\mu+n\over
2}-1,{\mu-n \over 2}-1)$, with $n$ an integer and $\mu$ a complex number. 
The representations are given by the following action
\begin{equation}\label{iii}
T_{n \mu}(g)f(z_1,z_2)=f(\alpha z_1 +\gamma z_2,\beta z_1 +\delta z_2).
\end{equation}
One simple realization of these functions is given by the functions of
one complex variables defined as \begin{equation} \phi(z)=f(z,1).
\end{equation}
On this set of functions the representation operators act in the
following way
\begin{equation}\label{actionn}
T_{n \mu}(g)\phi(z)=(\beta z +\delta)^{{\mu+n\over 2}-1}( {\bar \beta
\bar z +\bar \delta})^{{\mu-n\over 2}-1}\phi \left({\alpha z +\gamma
\over \beta z +\delta}\right).  
\end{equation} 
Two representations $T_{n_1\mu_2}(g)$ and $T_{n_1\mu_2}(g)$ are
equivalent if $n_1=-n_2$ and $\mu_1=-\mu_2$.

Unitary representations of $SL(2,C)$ are infinite dimensional.  They
are a subset of the previous ones corresponding to the two possible
cases: $\mu$ purely imaginary ($T_{n,i \rho}(g)$ $\mu=i\rho$,
$\rho=\bar \rho$, known as the {\em principal series}), and $n=0$,
$\mu=\bar \mu=\rho$, $\rho \neq 0$ and $-1 < \rho < 1$ ($T_{0
\rho}(g)$the {\em supplementary series}).  From now on we concentrate
on the principal series unitary representations $T_{n i\rho}(g)$ which
we denote simply as $T_{n \rho}(g)$ (dropping the $i$ in front of
$\rho$).  The invariant scalar product for the principal series is
given by \begin{equation} (\phi,\psi)= \int \bar\phi(z) \psi(z) dz,
\end{equation}
where $dz$ denotes $dRe(z) dIm(z)$.

There is a well defined measure on $SL(2,C)$ which is right-left
invariant and invariant under inversion (namely,
$dg=d(gg_0)=d(g_0g)=d(g^{-1})$).  Explicitly, in terms of the
components in (\ref{g})
\begin{eqnarray}\label{mea}
dg = \left( {i \over 2}\right)^3 {d\beta d\gamma d\delta \over
|\delta|^2} = \left( {i \over 2}\right)^3 {d\alpha d\gamma d\delta
\over |\gamma|^2} = \left( {i \over 2}\right)^3 {d\beta d\alpha
d\delta \over |\beta|^2} = \left( {i \over 2}\right)^3 {d\beta d\gamma
d\alpha \over |\alpha|^2},\end{eqnarray} where $d\alpha$, $d\beta$,
$d\gamma$, and $d\delta$ denote integration over the real and
imaginary part respectively.

Every square-integrable function, i.e, $f(g)$ such that
\begin{equation}
\int |f(g)|^2 dg \le \infty,
\end{equation} has a well defined Fourier transform defined as
\begin{equation}
F(n,\rho)=\int f(g) T_{n,\rho}(g) dg.
\end{equation}
This equation can be inverted to express $f(g)$ in terms of
$T_{n,\rho}(g)$.  This is known as the Plancherel theorem which
generalizes the Peter-Weyl theorem for finite dimensional unitary
irreducible representations of compact groups as $SU(2)$.  Namely,
every square-integrable function $f(g)$ can be written as
\begin{equation}\label{fu}
f(g)={1\over 8 \pi^4}\sum_n \int {\rm
Tr}[F(n,\rho)T_{n,\rho}(g^{-1})](n^2+\rho^2) d\rho, \end{equation}
where only components corresponding to the principal series are summed
over (not all unitary representations are needed)
\footnote{If the
function $f(g)$ is infinitely differentiable of compact support then
it can be shown that $F(n,\rho)$ is an analytic function of $\rho$ and
an expansion similar to (\ref{fu}) can be written in terms of
non-unitary representations.}
, and
\begin{equation}
{\rm Tr}[F(n,\rho)T_{n,\rho}(g^{-1})]=\int {\cal F}_{n\rho}(z_1,z_2)
{\cal T}_{n\rho}(z_2,z_1;g) dz_1dz_2.\end{equation} ${\cal
F}_{n\rho}(z_1,z_2)$, and ${\cal T}_{n\rho}(z_2,z_1;g)$ correspond to
the kernels of the Fourier transform and representation respectively
defined by their action on the space of functions $\phi(z)$ (they are
analogous to the momenta components and representation matrix elements
in the case of finite dimensional representations), namely
\begin{equation}
F(n,\rho)\phi(z):= \int f(g)T_{n\rho}(g)\phi(z) dg:=\int {\cal
F}_{n\rho}(z,\tilde z) \phi(\tilde z) d\tilde z, \end{equation} and
\begin{equation}
T_{n,\rho}(g)\phi(z):= \int {\cal
T}_{n\rho}(z,\tilde z;g) \phi(\tilde z) d\tilde z.\end{equation}
From (\ref{actionn}) we obtain that 
\begin{equation}
{\cal T}_{n\rho}(z,\tilde z;g)=(\beta z +\delta)^{{\rho+n\over 2}}( {\bar
\beta \bar z +\bar \delta})^{{\rho-n\over 2}} \delta\left(\tilde z-{\alpha z
+\gamma \over \beta z +\delta}\right).
\end{equation}
The resolution of the identity takes the form
\begin{equation}\label{fu1}
\delta(g)={1\over 8 \pi^4}\sum_n \int {\rm
Tr}[T_{n,\rho}(g)](n^2+\rho^2) \ d\rho.
\end{equation}

\subsubsection{The canonical basis}

There exists an alternative realization of the representations in terms of
the space of homogeneous functions $f(z_1,z_2)$ defined above\cite{ru}.
Because of homogeneity (\ref{hom}) any $f(z_1,z_2)$ is completely determined
by its values on the sphere $S^3$
\begin{equation}\label{up}
|z_1|^2+|z_2|^2=1.
\end{equation}
As it is well now there is an isomorphism between $S^3$ and $SU(2)$ given by
\begin{equation}
u=\left[\begin{array}{c}\bar z_2 \ -\bar z_1 \\ z_1 \ \ \ \ \, z_2 
\end{array}\right] \end{equation} for $u\in SU(2)$ and $z_i$ satisfying
(\ref{up}). Alternatively we can define the function $\phi(u)$ of $u\in
SU(2)$ as \begin{equation}
\phi(u):=f(u_{21},u_{22}),
\end{equation} with $f$ as in (\ref{hom}).
Due to (\ref{hom}) $\phi(u)$  has the following ``gauge'' behavior 
\begin{equation}\label{gauge}
\phi(\gamma u)= e^{i\omega(a-b)}\phi(u)= e^{i\omega n}\phi(u),
\end{equation}
for $\gamma= \left[ \begin{array}{c} e^{i \omega} \ 0 \\ 0 \ e^{-i
\omega} \end{array} \right]$.  The action of $T_{n \rho}(g)$ on
$\phi(u)$ is induced by its action on $f(z_1,z_2)$ (\ref{iii}).  We
can now use Peter-Weyl theorem to express $\phi(u)$ in terms of
irreducible representations $D^j_{q_1q_2}(u)$ of $SU(2)$. However, 
due to (\ref{gauge}) only the functions
$\phi^j_q(u)=(2j+1)^{\SC 1/2} D^j_{n q_2}(u)$ are needed (where
$j=|n|+k$, $k=0,1,\dots$). Therefore $\phi(u)$ can be written as
\begin{equation}
\phi(u)=\sum^{\infty}_{j=n} \sum^j_{q=-j} d^j_q \ \ \phi^j_q(u).
\end{equation} 
This set of functions is known as the canonical basis.  This basis is
better suited for generalizing the Euclidean spin
foam models, since the notation maintains a certain degree of
similarity with the one in \cite{dfkr,ac}.  We can use this basis to
write the matrix elements of the operators $T_{n,\rho}(g)$, namely
\begin{equation}\label{ma}
D^{n \rho}_{j_1 q_1 j_2 q_2}(g)=\int_{\SC SU(2)} \bar
\phi^{j_1}_{q_1}(u) \ \left[T_{n \rho}(g) \phi^{j_2}_{q_2}(u)\right]
du.  \end{equation} Since $T_{n_1 n_2}(u_0)\phi(u)=\phi(u u_0)$,
invariance of the $SU(2)$ Haar measure implies that
\begin{equation}\label{su}
D^{n \rho}_{j_1 q_1 j_2 q_2}(u_0)=\delta_{j_1 j_2} \ D^{j_1}_{q_1 q_2}(u_0).
\end{equation} In terms of these matrix elements equation (\ref{fu})
acquires the more familiar form
\begin{equation}\label{pass}
f(g)=\sum^{\infty}_{n=0} \int^{\infty}_{\rho=0}
\left[\sum^{\infty}_{j_1,j_2=n} \sum^{j_1}_{q_1=-j_1}
\sum^{j_2}_{q_2=-j_2} \bar D^{n,\rho}_{j_1 q_1 j_2 q_2}(g) f^{j_1 q_1
j_2 q_2}_{n,\rho} \right](n^2+\rho^2) d\rho, \end{equation} where
\begin{equation}\label{furie}
f^{j_1 q_1 j_2 q_2}_{n,\rho}=\int f(g) D^{n,\rho}_{j_1 q_1 j_2 q_2}(g) dg,
\end{equation}
and the quantity in brackets
represents the trace in (\ref{fu}). In the same way we can translate equation
(\ref{fu1}) obtaining
\begin{equation}\label{vani}
\delta(g)=\sum^{\infty}_{n=0} \int^{\infty}_{\rho=0}
\left[\sum^{\infty}_{j=n} \sum^{j}_{q=-j} \bar D^{n,\rho}_{j q j q}(g) 
\right](n^2+\rho^2) d\rho =\sum^{\infty}_{n=0} \int^{\infty}_{\rho=0}
{\rm Tr}\left[\bar D^{n,\rho}(g)\right](n^2+\rho^2) d\rho.
\end{equation}  
Using equations (\ref{ma}) and (\ref{su}), we can compute
\begin{eqnarray}\label{proj}
\int_{\SC SU(2)}\! \! \!  D^{n,\rho}_{j_1 q_1 j_2 q_2}(u)\ du = 
\delta_{j j_2}\int_{\SC
SU(2)} D^{j}_{q  q_2}(u) du 
 =  \delta_{j_2 0} \delta_{j_1 0}.
\end{eqnarray} 

\subsubsection{On the tensor product of two irreducible representations}

The tensor product of two irreducible representations of the principal series 
$T_{n_1 \rho_1}$ and $T_{n_2 \rho_2}$ decomposes into a direct integral of 
irreducible representations $T_{n \rho}$ for those $n$'s
such that $n+n_1+n_2$ is an even integer and no restriction for $\rho$. 
For a proof of this assertion, and for explicit realizations of the 
tensor product of two representations of the 
principal series see \cite{nai}.

\subsubsection{Generators and Casimir operators}

An infinitesimal $g \in SL(2,C)$ in the adjoint representation can 
be parametrized by the six numbers $\lambda_{\mu \nu}=-\lambda_{\nu \mu}$ as
\begin{equation}
g = e + i \lambda_{\mu \nu} L^{\mu \nu}+ O(\lambda^2),
\end{equation}
where $i L^{\mu \nu} \in sl(2,c)$, the algebra of $SL(2,C)$.
The corresponding irreducible representation operator $T^{n \rho}(g)$
has the form
\begin{equation}
T_{n \rho}(g) = \hat 1 + i \lambda_{\mu \nu} {\hat L_{\SC (n
\rho)}}^{\mu \nu}+ O(\lambda^2).
\end{equation}
There are two Casimir operators in $SL(2,C)$ corresponding to $L_{\mu \nu}
L^{\mu \nu}$ and $L_{\mu \nu} {}^*L^{\mu \nu}$ respectively, namely 
\begin{equation}\label{c1}
\hat{\cal C}_{1{\SC(n,\rho)}} =\hat L_{{\SC (n \rho)}\mu \nu} \hat
L_{\SC (n \rho)}^{\mu \nu}= {1 \over 4} (n^2-\rho^2-4) \ \hat 1,
\end{equation} and \begin{equation}\label{c2}
\hat{\cal C}_{2 {\SC(n,\rho)}}=\epsilon_{\mu \nu \alpha \beta} \hat 
L_{{\SC (n \rho)}}^{\mu \nu} \hat 
L_{\SC (n \rho)}^{\alpha \beta}={1 \over 4} n \rho \
\hat 1. 
\end{equation}
 
\end{appendix}


\begin{thebibliography}{10}

\bibitem{BC2} JW Barrett, L Crane, {\em A Lorentzian Signature Model
for Quantum General Relativity\/}, gr-qc/9904025, Class Quant Grav 17
(2000) 3101-3118.

\bibitem{lac} A Perez, C Rovelli, {\em Spin foam model for Lorentzian
General Relativity}, gr-qc/0009021.
    
\bibitem{Reisenberger} M Reisenberger, {\em Worldsheet formulations of
gauge theories and gravity}, talk given at the 7th Marcel Grossmann
Meeting Stanford, July 1994; gr-qc/9412035.

\bibitem{Iwasaki} J Iwasaki, {\em A definition of the Ponzano-Regge
quantum gravity model in terms of surfaces}, J Math Phys 36 (1995)
6288.

\bibitem{Baez} J Baez, {\em Spin Foam Models}, Class  Quant  Grav
15 (1998) 1827-1858; gr-qc/9709052.  {\em An Introduction to Spin Foam
Models of Quantum Gravity and BF Theory}, to appear in to appear in
``Geometry and Quantum Physics'', eds Helmut Gausterer and Harald
Grosse, Lecture Notes in Physics (Springer-Verlag, Berlin);
gr-qc/9905087.

\bibitem{rr} M Reisenberger, C Rovelli, {\em Sum over Surfaces form of
Loop Quantum Gravity}, Phys  Rev  D56 (1997) 3490-3508.  C Rovelli,
{\em Quantum gravity as a sum over surfaces}, Nucl  Phys  B57 (1997)
28-43.  C Rovelli, {\em The projector on physical states in loop
quantum gravity}, gr-qc/9806121.

\bibitem{Roberto} R De Pietri, {\em Canonical Loop Quantum
Gravity and Spin Foam Models}, Proceeding of the XXIII SIGRAV
conference, Monopoli (Italy), September 1998. 

\bibitem{KF} L Freidel, K Krasnov, {\em Spin Foam Models and the
Classical Action Principle}, Adv Theor Math Phys 2 (1999) 1183-1247, 
hep-th/9807092. 

\bibitem{BC}
JW Barrett, L Crane, {\em Relativistic spin networks and
quantum gravity}, J Math Phys 39 (1998) 3296.

\bibitem{misner} C Misner, {\em Feynman quantization of General
Relativity}, Rev  Mod  Phys  29 (1957) 497.

\bibitem{haw} SW Hawking, {\em The Path-Integral Approach to
Quantum Gravity}, in ``General Relativity: An Einstein
Centenary Survey'', SW Hawking and W Israel eds (Cambridge
University Press, Cambridge 1979).

\bibitem{Reisenberg97} M Reisenberger, {\em A left-handed simplicial
action for Euclidean general relativity}, {Class Quantum Grav} {14}
(1997) 1730--1770; gr-qc/9609002; gr-qc/9711052; gr-qc/9903112

\bibitem{iwa0} J Iwasaki, {\em A surface theoretic model of quantum
gravity}, gr-qc/9903112.  J Iwasaki, {\em A lattice quantum gravity
model with surface-like excitations in 4-dimensional spacetime},
gr-qc/0006088.

\bibitem{ac} A Perez, C Rovelli, {\em A spin foam model without bubble
divergences}, gr-qc/0006107.

\bibitem{cuate} J F Plebanski,{\em On the separation of Einsteinian
substructures}, J Math Phys 12, (1977) 2511.

\bibitem{dfkr} R De Pietri, L Freidel, K Krasnov, C Rovelli,
{\em Barrett-Crane model from a Boulatov-Ooguri field theory
over a homogeneous space}, Nuclear Physics, to appear,
hep-th/9907154.

\bibitem{cm} M Reisenberger, C Rovelli, {\em Spinfoam models as
Feynman diagrams}, gr-qc/0002083.  M Reisenberger, C Rovelli, {\em
Spacetime as a Feynman diagram: the connection formulation},
gr-qc/0002095.

\bibitem{t}
V Turaev, {\em Quantum invariants of 3-manifolds and a glimpse of
shadow topology} in {\em Quantum Groups}, Springer Lecture Notes in
Mathematics 1510, pp 363-366 (Springer-Verlag, New York, 1992); {
``Quantum Invariants of Knots and 3-Manifolds''} (de Gruyter, New York,
1994).

\bibitem{Ooguri:1992b} H Ooguri, {\em Topological Lattice
Models in Four Dimensions}, Mod Phys Lett {A7} (1992) 2799.

\bibitem{CraneYetter} L Crane and D Yetter, {\em A Categorical
construction of 4-D topological quantum field theories}, in 
``Quantum Topology'', L Kaufmann and R Baadhio eds (World Scientific,
Singapore 1993); hep-th/9301062.

\bibitem{CraneYetter1} L Crane, L Kauffman and D Yetter, {\em
State-Sum Invariants of 4-Manifolds}, J Knot Theor Ramifications { 6}
(1997) 177--234; hep-th/9409167.

\bibitem{area} C Rovelli, L Smolin, {\em Discreteness of the area and
volume in quantum gravity}, Nucl Phys B442 (1995), 593-622.  Erratum:
Nucl Phys B456 (1995), 734.

\bibitem{fac} A Perez, Carlo Rovelli, {\em Finiteness of a spinfoam
model for euclidean quantum general relativity}, very soon on gr-qc. 

\bibitem{BB} J Baez, J W Barret, {\em Integrability for relativistic
spin networks}, in preparation.

\bibitem{barret} JW Barrett, {\em The classical evaluation of
relativistic spin networks}, Adv Theor Math Phys 2 (1998) 593-600.

\bibitem{gel} IM Gel'fand, MI Graev, N Ya Vilenkin, ``Generalized
Functions'', volume 5, Integral Geometry and Representation Theory,
(Academic Press, New York 1966).

\bibitem{ru} W Ruhl, ``The Lorentz Group and Harmonic Analysis''  
(WA Benjamin Inc, New York 1970).

\bibitem{ted} A Held, ET Newman, R Posadas, {\em The Lorentz Group
and the Sphere}, J Math Phys 11 (1970) 3145.

\bibitem{nai} MA Naimark, {\em Decomposition of a tensor product of irreducible 
representations of the proper Lorentz group into irreducible representations.}, 
Amer Math Soc Translations ser 2 36 101-136 (1964).

\bibitem{chor} D Oriti, R M Williams, {\em Gluing 4-simplices: a
derivation of the Barrett-Crane spin foam model for Euclidean quantum
gravity}, gr-qc/0010031.

\end{thebibliography}
\end{document}